\documentclass[aps,twocolumn,showpacs,preprintnumbers,nofootinbib,prd,superscriptaddress,groupedaddress,10pt]{revtex4-1}

\usepackage{graphicx,amssymb,amsmath,amsthm,amsfonts,mathrsfs,dsfont,epsfig,epsf}
\usepackage[hidelinks]{hyperref}
\usepackage[usenames]{color}
\usepackage{epstopdf}
\usepackage{aas_macros}
\usepackage{bm}
\usepackage{dcolumn}
\usepackage[latin1]{inputenc}
\usepackage{latexsym}
\usepackage{rotating}
\usepackage{longtable}
\usepackage{enumerate}
\usepackage{tensor,multirow}
\usepackage{url,float}
\def\be{\begin{equation}}
\def\ee{\end{equation}}
\def\beq{\begin{eqnarray}}
\def\eeq{\end{eqnarray}}

\newcommand{\ben}{\begin{enumerate}}
\newcommand{\een}{\end{enumerate}}
\newcommand{\bi}{\begin{itemize}}
\newcommand{\ei}{\end{itemize}}

\def\ga{\mathrel{\raise.3ex\hbox{$>$\kern-.75em\lower1ex\hbox{$\sim$}}}}
\def\la{\mathrel{\raise.3ex\hbox{$<$\kern-.75em\lower1ex\hbox{$\sim$}}}}

\def\I_M{{I_{\scriptscriptstyle M\times M}}}

\begin{document}

\title{Dynamical friction in slab geometries and accretion disks}

\author{Rodrigo Vicente$^{1}$, Vitor Cardoso$^{1,2}$, Miguel Zilh\~ao$^{1}$}

\affiliation{$^1$CENTRA, Departamento de F\'isica, Instituto Superior
T\'ecnico, Universidade de Lisboa, Avenida Rovisco Pais 1,
1049 Lisboa, Portugal} 
\affiliation{$^2$Theoretical Physics Department, CERN 1 Esplanade des Particules, Geneva 23, CH-1211, Switzerland}

\begin{abstract}
The evolution of planets, stars and even galaxies is driven, to a large extent, by dynamical friction of gravitational origin. 
There is now a good understanding of the friction produced by extended media, either collisionless of fluid-like.
However, the physics of accretion or protoplanetary disks, for instance, is described by slab-like geometries instead,
compact in one spatial direction. Here, we find, for the first time, the gravitational wake due to a massive perturber moving through a slab-like medium, describing e.g. accretion disks with sharp transitions. We show that dynamical friction in such environments can be substantially {\it reduced}
relatively to spatially extended profiles. Finally, we provide simple and accurate expressions for the gravitational drag force felt by the perturber, in both the subsonic and supersonic regime.
\end{abstract}
\keywords{Gravitational drag --- accretion disks --- protoplanetary disks}
\maketitle

\section{Introduction} 
Drag forces of electromagnetic origin are ubiquitous in everyday life, and shape -- to some extent -- our own civilization.
On large scales, such as those of stars and galaxies, gravitational drag forces dominate the dynamics. When stars or planets move through a medium, a wake of fluctuation in the medium density is left behind. Gravitational drag (also known as dynamical friction) is caused by the backreaction of the wake on the moving object. Gravitational drag determines a number of features of astrophysical systems, for example planetary migration within disks, the sinking of supermassive black holes to the center of galaxies
or the motion of stars within galaxies on long timescales~\citep{Chandrasekhar:1943ys,1943ApJ....97..263C,1943ApJ....98...54C,Ostriker:1998fa,binney2011galactic}.

Dynamical friction is well studied when the object moves in an infinite (collisionless or fluid-like) medium~\citep{Chandrasekhar:1943ys,Ostriker:1998fa,Kim:2007zb,Kim:2008ab,1999ApJ...522L..35S}. Most of the rigorous treatments of dynamical friction in the literature 
-- with a few exceptions such as Ref.~\cite{namouni,Muto:2011qv,canto} -- consider as a setup an infinite three-dimensional medium. Clearly, such idealization breaks down in thin accretion or protoplanetary disks, where the geometry of the problem is more ``slab-like''~\citep{Novikov:1973kta,2011ARA&A..49..195A}. In this context, Ref.~\cite{Muto:2011qv} obtained estimates for the dynamical friction under the assumption of a steady state, and using a two-dimensional approximation for the gaseous medium. However, as the authors point out, their simplified approach has some limitations, and a fully three-dimensional treatment is needed (see Sec.~\ref{sec:conclusions} for further details). Also assuming a steady state, Ref.~\cite{canto} computed the gravitational drag on a hypersonic perturber moving in the midplane of a gaseous disk with Gaussian vertical density stratification. However, they did not studied how (and if) this steady state is dynamically attained.

In this work we compute, for the first time, the gravitational wake produced and the time-dependent force felt by the massive perturber moving in a three-dimensional medium with a slab-like geometry, subjected to either Dirichlet or Neumann conditions at the boundaries. 
This setup is a more faithful approximation to the physics of thin disks and we expect some of our main findings to carry over, at least at the qualitative level, to more generic physical situations where boundaries play a role.

\section{3D Gravitational drag}
The linearized equations for the perturbed density $\rho \equiv \rho_0[1+\alpha(\textbf{r},t)]$ and velocity $ \textbf{v} \equiv c \boldsymbol{\beta}(\textbf{r},t)$ of an adiabatic gaseous medium which is under the influence of an external potential $\phi_{\text{ext}}(\textbf{r},t)$ are~\citep{Ostriker:1998fa}
\beq
&&\frac{1}{c}\frac{\partial}{\partial t}\alpha +\boldsymbol{\nabla}\cdot \boldsymbol{\beta}=0 \,, \label{linearizedeq1}\\ 
&&\frac{1}{c} \frac{\partial}{\partial t}\boldsymbol{\beta}+\boldsymbol{\nabla}\alpha=-\frac{1}{c^2}\boldsymbol{\nabla}\phi_{\text{ext}}\,, \label{linearizedeq2}
\eeq
with $c$ the sound speed on the unperturbed medium, and $\alpha,|\boldsymbol{\beta}|\ll 1$. These equations can be combined to obtain the inhomogeneous wave equation 
\begin{equation} 
	\nabla^2 \alpha-\frac{1}{c^2}\frac{\partial^2 \alpha}{\partial t^2}=-\frac{1}{c^2}\nabla^2\phi_{\text{ext}}\,.\label{densdiffeq}
\end{equation}  
If the external influence is due to the gravitational interaction with a massive perturber of mass density $\rho_{\text{ext}}(\textbf{r},t)$,
\begin{equation}
\nabla^2\phi_{\text{ext}}=4 \pi \mathcal{G} \rho_{\text{ext}}\, ,\label{eq:poisson}
\end{equation}
where $\mathcal{G}$ is the gravitational constant.
Equation~\eqref{densdiffeq} can be solved employing the method of Green's function. The Green's function $G(\textbf{r},t|\textbf{r}',t')$ of the differential operator in the left-hand side of Eq.~\eqref{densdiffeq} satisfies
\begin{equation}
\nabla^2 G-\frac{1}{c^2}\frac{\partial^2 G}{\partial t^2}=- \delta^{(3)}(\textbf{r}-\textbf{r}') \delta(t-t')\,.\label{flat_green}
\end{equation}

The problem of finding, at linear order, the perturbed density $\rho(\textbf{r},t)$ of an infinite three-dimensional gaseous medium, due to the gravitational pull of a point-like mass $M$ moving at velocity $\textbf{v}$, was solved in Ref.~\cite{Ostriker:1998fa}. 
The dynamical friction felt by the moving mass was therein computed to be
\beq
F&=&\frac{ (\mathcal{G} M)^2 \rho_0}{c^2} I(\mathcal{M},t)\,,\label{3Ddragforce}\\
I&=&-\frac{4 \pi}{\mathcal{M}^2}\left[\frac{1}{2} \log\left(\frac{1+\mathcal{M}}{1-\mathcal{M}}\right)- \mathcal{M}\right]\,,\quad \mathcal{M}<1\label{subsonic_ostriker}\\
I&=&-\frac{4 \pi}{\mathcal{M}^2} \left[\frac{1}{2} \log\left(1-\frac{1}{\mathcal{M}^2}\right)+ \log\left(\frac{\mathcal{M} c t}{r_\text{min}}\right)\right]\,, \mathcal{M}>1\nonumber\\
\label{supersonic_ostriker}
\eeq
with $\mathcal{M}\equiv v/c$ the Mach number, $r_\text{min}$ the effective size of the perturber, and assuming that $r_\text{min}<(\mathcal{M}-1) c t$.\footnote{In fact, the gravitational drag force felt by a supersonic point-like mass is infinite. Thus, a cutoff $r_\text{min}$ is needed (\textit{i.e.}, the perturber must have finite size).}
Notice that the dynamical friction always opposes the perturber's motion (\textit{i.e.}, $F<0$).

\section{Gravitational drag in slab geometries}
Consider now a medium with a slab-like profile: of constant density and extending to arbitrarily large spatial distances in the $x$ and $y$ directions, but of compact support in the $z$-direction, with thickness $2 L$ for  $-L\leq z \leq L$.
The linear perturbation in the pressure is $\delta p= (\partial p/\partial \rho) \delta \rho=c^2 \rho_0 \alpha(\textbf{r},t)$. Thus, the physically relevant setup $\delta p(z=-L,t)=\delta p(z=L,t)=0$ corresponds to Dirichlet conditions on $\alpha(\textbf{r},t)$ at the boundaries of the slab. For completeness, we provide results for Neumann conditions as well.

Define $T\equiv t-t'$ and $\textbf{R} \equiv(x-x',y-y')$.
The solution of Eq.~\eqref{flat_green} satisfying Dirichlet boundary conditions is
\beq
G&=&\sum_{n=0}^{+\infty} \frac{c}{2 \pi L} \frac{\cos\left(m_n D\right)}{D} \sin[m_n (z+L)] \sin[m_n (z'+L)]\nonumber\\
&&\times \Theta(c T - R)\,,
\eeq
with $D\equiv \sqrt{c^2 T^2-R^2}$ and $m_n\equiv n \pi/(2L)$. After some algebra this expression can be put in the form
\beq
G&=&\sum_{l=-\infty}^{+\infty} \frac{(-1)^l}{4\pi c^2 \sqrt{(z-(-1)^l z'-2 l L)^2+R^2}}\nonumber\\
&&\times  \delta\left(\frac{\sqrt{(z-(-1)^l z'-2 l L)^2+R^2}}{c}-T\right)\,.	
\eeq

The gravitational interaction between the medium and an external massive perturber is governed by Poisson's equation
\eqref{eq:poisson}. Thus, we find the solution to Eq.~\eqref{densdiffeq} with Dirichlet boundary conditions:
\beq
\alpha&=&\sum_{l=-\infty}^{+\infty}\int d^3\textbf{r}' dt'\frac{(-1)^l \mathcal{G} \rho_\text{ext}(\textbf{r}',t')}{c^2\sqrt{(z-(-1)^l z'-2 l L)^2+R^2}}\nonumber\\
&&\times \delta\left(\frac{\sqrt{(z-(-1)^l z'-2 l L)^2+R^2}}{c}-T\right)\,.
\eeq

The index $l$ has an important physical meaning: it is the number of ``reflections'' that fluctuations have undergone at the boundaries. Therefore, the density $\alpha$ is expanded in terms of the number of \textit{echoes} of the Green's function on the slab. This result is analogous to that of a signal propagating along a four-dimensional brane in a five-dimensional Kaluza-Klein spacetime, except that boundary conditions are of the Neumann type for that problem~\citep{Barvinsky:2003jf}. Notice also that the $l=0$ term in the expansion corresponds to direct propagation, and describes the fluctuations not sensitive to the boundaries. Not surprisingly, this term describes exactly the solution for a three-dimensional infinite medium~\citep{Ostriker:1998fa}.

Now consider a particle of mass $M$ moving with velocity $\textbf{v}$ on a straight-line through the medium, describing the trajectory $\textbf{r}(t)=(x(t),y(t),z(t))=(v t,0,0)$, with $v>0$. This perturbation is turned on at $t=0$, with $\rho_{\text{ext}}=M \delta(x-v t)\delta(y)\delta(z) \Theta(t)$. Under these conditions, the perturbation in the medium density is
\beq
&&\alpha=\frac{\mathcal{G} M}{c^2}\sum_{l=-\infty}^{+\infty} (-1)^{\eta l}\int_{-\infty}^{+\infty} dw  \,\Theta(w+x)\nonumber\\
&\times&\frac{\delta\left(w+s+\mathcal{M} \sqrt{(z-2 l L)^2+w^2+y^2} \right)}{\sqrt{(z-2 l L)^2+w^2+y^2}}  \,,\label{slabdensity}
\eeq
with $w \equiv x'-x$, $s \equiv x-v t$, and $\eta \equiv 1,\,0$ for Dirichlet and Neumann conditions, respectively.

\subsection{Subsonic perturbers}
%
\begin{figure}[]
\includegraphics[width=1\linewidth]{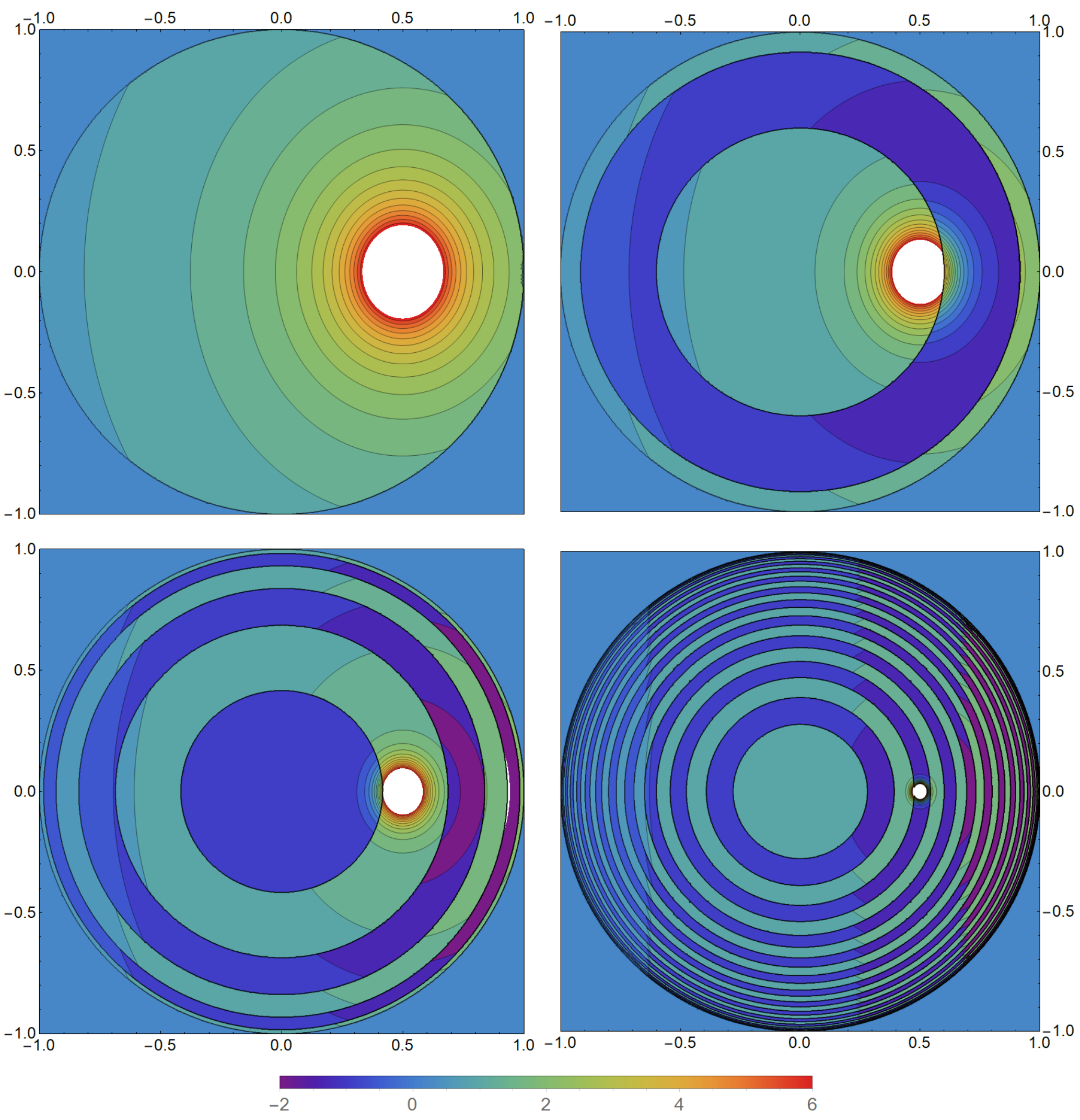}
\caption{Density perturbation $\alpha c^3 t/(\mathcal{G} M)$ in a gaseous slab, along $z=0$, due to the gravitational interaction with a subsonic perturber with Mach number $\mathcal{M}=0.5$, for $c t/L=(0.5,5,11,50)$ (left to right, top to bottom). The horizontal axis represents the coordinate $x/(c t)$, and the vertical axis the coordinate $y/(c t)$. The contours represent curves of constant density. The observed ripples centered at the origin -- which turn on at $c t/L\geq1$, but are only seen in $z=0$ at $ct/L\geq2$ -- are echoes of the original density fluctuation. Each ripple is associated with a different $l$-term.}
\label{fig:slabsubsonicz0D}
\end{figure}
Consider first perturbers with Mach number $\mathcal{M}<1$. The argument of the delta in Eq.~\ref{slabdensity} then vanishes for
\begin{equation}
w=w_l\equiv -\frac{s+\mathcal{M}\sqrt{s^2+(1-\mathcal{M}^2)\left[y^2+(z-2 l L)^2\right]}}{1-\mathcal{M}^2}\,.\nonumber
\end{equation}  
Each $l$-term contribution to $\alpha(\textbf{r},t)$ vanishes for $-x>w_l$, or equivalently,
\begin{equation}
x^2+y^2+(z-2 l L)^2>c^2 t^2\,.
\end{equation}
This is a manifestation of the causality principle. The perturber is turned on at $(t,x,y,z)=0$ and moves with $v<c$. The fluctuation, on the other hand, propagates with a speed $c$. These two facts imply that at instant $t$, the maximum region of influence of the massive particle is the region in the slab defined by $x^2+y^2+z^2\leq c^2t^2$. This is the ``region of influence'' of the $l=0$ term. At fixed $t$, each $l$-term has a different region of influence. Larger $l$'s probe smaller regions since the fluctuation is busy traveling between the boundaries and is unable to probe large $x,y$ directions.

\begin{figure}[]
	\includegraphics[width=1\linewidth]{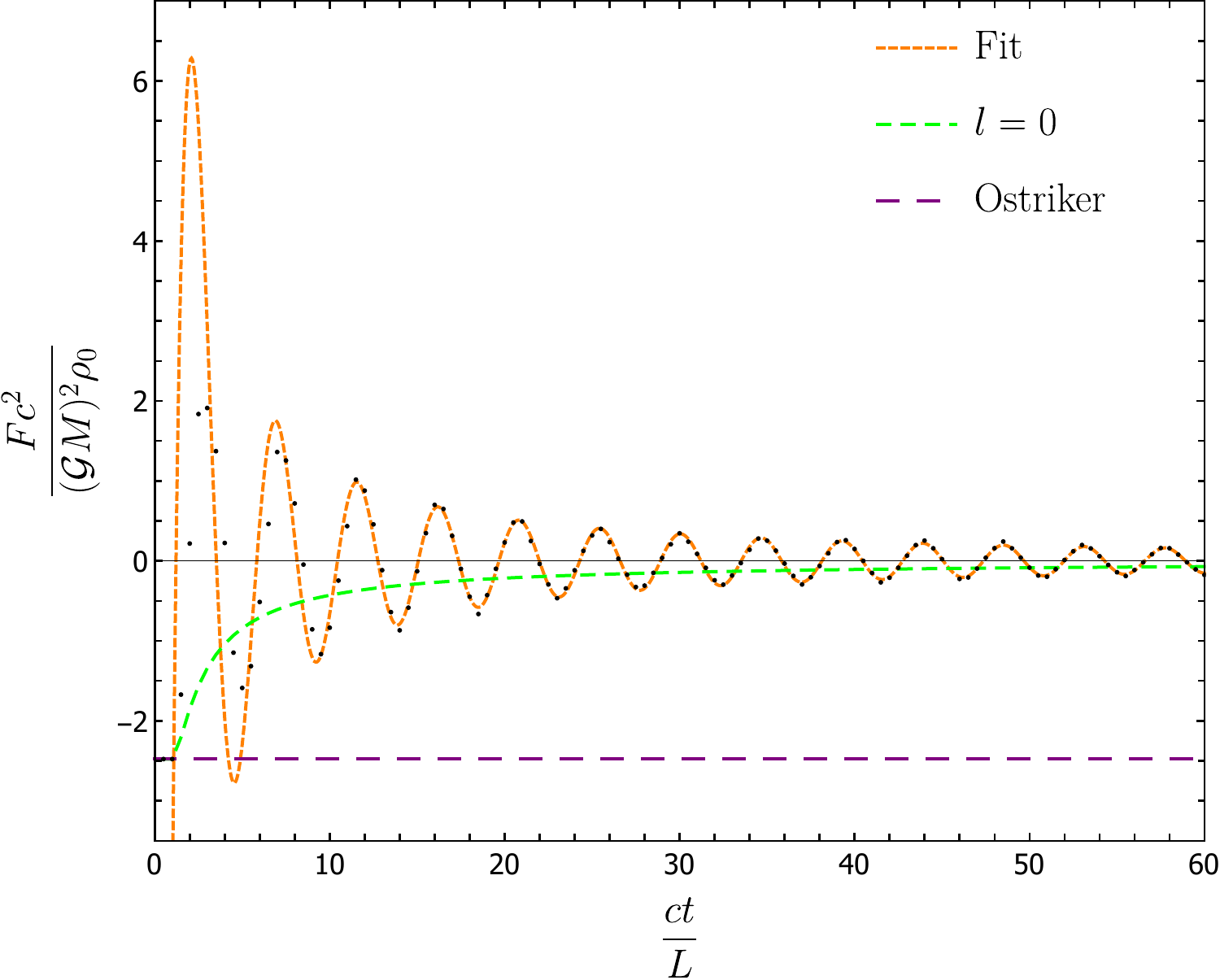} 
	\caption{Gravitational drag force $F c^2/\left((\mathcal{G}M)^2 \rho_0\right)$ felt by a particle moving at a subsonic Mach number $\mathcal{M}=0.5$ as function of $c t/L$ (black dots). The results are in agreement with the predicted early- and late-time behavior of the force, as described by Eqs.~\eqref{slabDragforceearly}and~\eqref{slabDragforcelateD}, respectively. Notice that the early-time force ($c t/L<1$) is independent of the boundary conditions, and, thus, is the same as for non-compact geometries; therefore it is described by well-known results~\citep{Ostriker:1998fa} (purple dashed curve). At late times, the force oscillates with a period $\sim 4 L/c \exp\left( \dfrac{\mathcal{M}^{2.18}}{2 (1-\mathcal{M})^{0.32}}\right)$, and decays as $\sim L/(c t)$; the orange dashed curve is the fit expression~\eqref{fitsubD}. In green, we show the ($l=0$) contribution from the non-reflected wake.
		\label{fig:slabsubsonicforcetD}}
\end{figure}

\begin{figure}[]
	\includegraphics[width=1\linewidth]{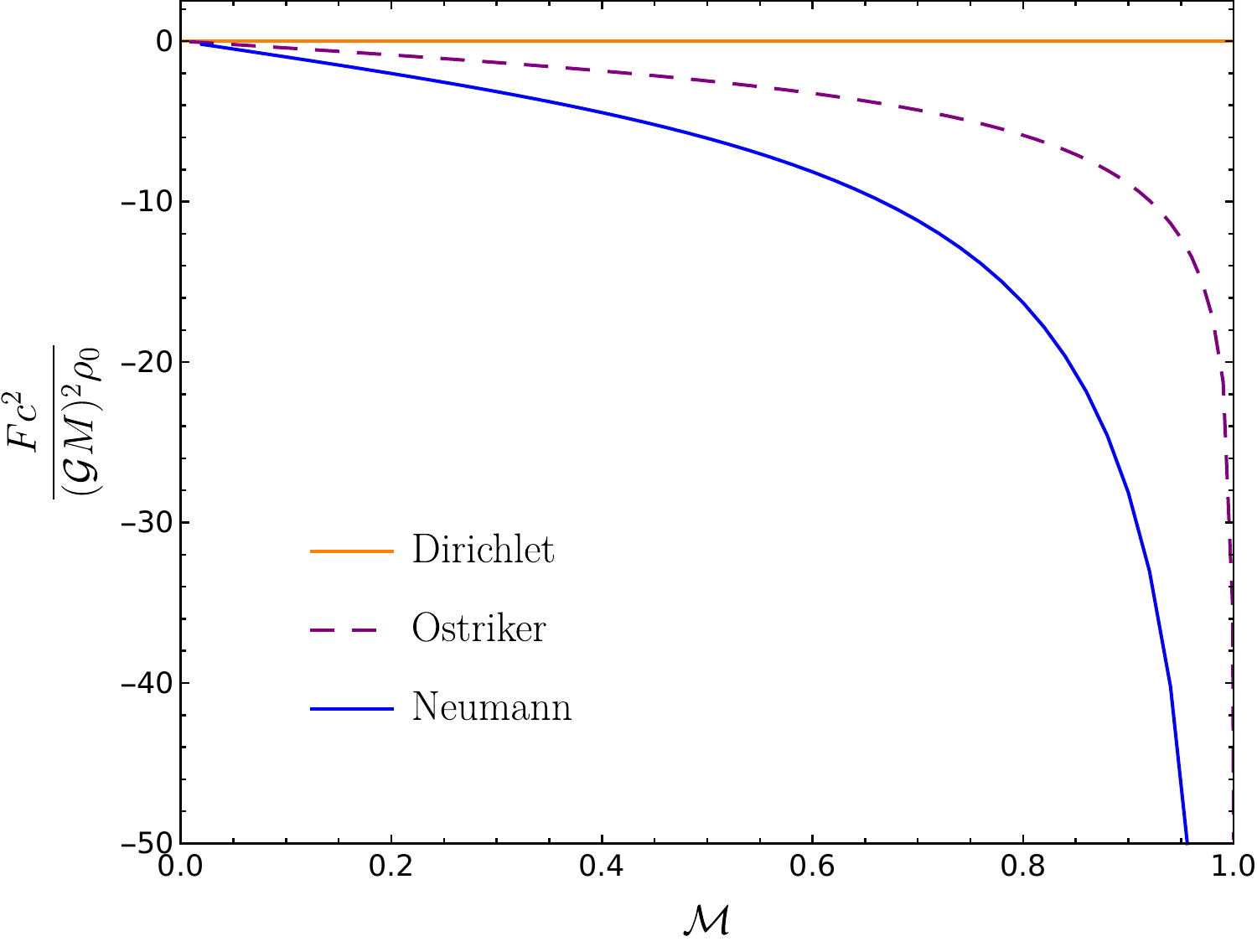} 
	\caption{Comparison between the early- [Eq.~\eqref{slabDragforceearly}] and late-time [Dirichlet: Eq.~\eqref{slabDragforcelateD}; Neumann: Eq.~\eqref{dragforceapplate}] drag force $F c^2/\left((\mathcal{G}M)^2 \rho_0\right)$ as function of the
		Mach number $\mathcal{M}$. In the subsonic regime, the dynamical friction due to a three-dimensional slab medium with Dirichlet (Neumann) conditions is always smaller (larger) in magnitude than the one due to an infinite three-dimensional medium.}
	\label{fig:slabsubsonicforcem}
\end{figure}

Notice that not all $l$-terms contribute to $\alpha$ at instant $t$. A given $l$ mode only contributes from $t_l=(2|l|-1)L/c$ onwards. 
Physically, this is due to these terms being echoes, and requiring therefore a finite time to reach the slab boundaries.
The only exception is the $l=0$ term, which contributes from $t=0$ onwards.

In summary, a massive particle moving at subsonic speeds through a gaseous slab causes a fluctuation
\begin{equation}
\alpha(\textbf{r},t)=\frac{\mathcal{G} M}{c^2}\sum_{l=-\infty}^{+\infty}  \frac{(-1)^{ \eta l}\Theta \left[c^2 t^2-x^2-y^2-(z-2 l L)^2\right]}{\sqrt{s^2+(1-\mathcal{M}^2)\left[y^2+(z- 2l L)^2\right]}} \,,\nonumber
\end{equation}
in the medium density, where we used the property $|A|\delta \left[A(w-w_l) \right]=\delta(w-w_l)$.
A contour plot of the density profile is shown in Fig.~\ref{fig:slabsubsonicz0D}, at different instants. The perturber is moving at a subsonic speed with Mach number $\mathcal{M}=0.5$. The results for $ct/L=0.5$ coincide {\it exactly} with the ones obtained for non-compact geometries~\citep{Ostriker:1998fa}, since the perturbation did not yet have time to reach the boundaries.

Let us now calculate the gravitational drag force felt by the moving particle. An infinitesimal element of the medium $\rho dx dy dz$ at $\textbf{r}$ acts gravitationally on a particle of mass $M$ (at position $\textbf{v} t$) through
\begin{equation}
d\textbf{F}(\textbf{r},t)= dx dy dz\frac{\mathcal{G} \rho M}{\left[(x-v t)^2+y^2+z^2\right]^{3/2}} (\textbf{r}-\textbf{v} t) \,.\label{force_drag}
\end{equation} 
By symmetry, the net force felt by the particle points in the $\textbf{x}$-direction. 
For times such that $c t/L<1$, the only contributing term is the $l=0$, and the force reduces to
\beq 
F&=& \frac{(\mathcal{G} M)^2 \rho_0}{c^2}   \int d\bar{z} d\bar{x} d\bar{y} \, \frac{\Theta \left[1-\bar{x}^2-\bar{y}^2-\bar{z}^2\right]}{\sqrt{(\bar{x}-\mathcal{M})^2+(1-\mathcal{M}^2)(\bar{y}^2+\bar{z}^2)}}\nonumber\\
&\times&\frac{\bar{x}-\mathcal{M}}{\left[(\bar{x}-\mathcal{M})^2+\bar{y}^2+\bar{z}^2 \right]^{3/2}}\,,	\label{slabDragforceearly}
\eeq
for both Dirichlet and Neumann conditions, where we defined barred coordinates $\bar{x}^i\equiv x^i/(c t)$. This expression is clearly time-independent, and the integration gives Eq.~\eqref{subsonic_ostriker}. Thus (not surprisingly) for $c t/L<1$ the perturbation did not yet probe the boundary and one recovers well-known results~\citep{Ostriker:1998fa}.

To find the force at late times $c t/L \gg 1$, we first break the expansion in even and odd $l$-terms, and define $l_1\equiv 2 l $ and $l_ 2\equiv 2l+1$, then, the drag force reads
\beq
F&\simeq& \frac{2 L}{c t} \frac{(\mathcal{G} M)^2 \rho_0}{c^2} \sum_{|l_1|\leq \text{int}\left[c t/(4 L)\right]} \int d\bar{x} d\bar{y} \frac{\bar{x}-\mathcal{M}}{\left[(\bar{x}-\mathcal{M})^2+\bar{y}^2\right]^{3/2}} \nonumber\,\, \\
&\times&\frac{\Theta \left[1-\bar{x}^2-\bar{y}^2-\left(\frac{4 L}{c t}\right)^2 l_1^2\right]}{\sqrt{(\bar{x}-\mathcal{M})^2+(1-\mathcal{M}^2)\left[\bar{y}^2+ \left(\frac{4 L}{c t}\right)^2 l_1^2\right]}} -(l_1 \to l_2)\nonumber\\
&=&0\,,\label{slabDragforcelateD}
\eeq
with $\text{int}(k)$ the integer part of $k$. Thus, we obtain the remarkable result that in slab geometries with Dirichlet boundary conditions there is no drag force at late times!

The numerical results of the integration of Eq.~\eqref{force_drag} are shown in Fig.~\ref{fig:slabsubsonicforcetD} for a fixed Mach number $\mathcal{M}=0.5$. The force is initially the same as that in extended geometries, Eq.~\eqref{subsonic_ostriker}.
However, after the fluctuations reach the boundary, such force changes.

It is amusing to see that for some time intervals the drag force acting on the perturber is {\it positive}. This can be traced by to the existence of regions of negative density fluctuation $\alpha$, which effectively act in a repulsive way on the particle, due to the deficit of matter in such region. Positive drag (sometimes called slingshot effect) does not arise with Neumann conditions, nor for an infinite three-dimensional medium, but nothing forbids it from appearing (and in fact it does, in slab geometries).

At late times we find a damped oscillatory behavior well described by (see also Fig.~\ref{fig:slabsubsonicforcetD})
\beq \label{fitsubD}
F &\simeq& \frac{(\mathcal{G} M)^2 \rho_0}{c^2} \frac{\mathcal{A}}{(c t/L)^\mathcal{B}}  \cos\left(\frac{2 \pi}{\mathcal{T}} \frac{c t}{L}+ \varphi \right)\,,
\eeq
where $\mathcal{A}$, $\mathcal{B}$, $\mathcal{T}$, and $\varphi$ are functions of $\mathcal{M}$. The power $\mathcal{B}$ is $\sim 0.6 (\pm 0.2)$ for $\mathcal{M}\leq 0.3$, and $\sim 1 (\pm 0.2)$ for $\mathcal{M}> 0.3$. The period of oscillation follows the law
\begin{equation*}
	\mathcal{T} \simeq 4 L/c \exp\left( \dfrac{\mathcal{M}^{2.18}}{2 (1-\mathcal{M})^{0.32}}\right)\,.
\end{equation*}

If Neumann conditions were used instead, our numerical results show that, at late times $c t/L\gg 1$, the moving particle feels a steady drag force well described by 
\begin{equation} 
F \simeq - 7.864 \frac{(\mathcal{G} M)^2 \rho_0}{c^2} \frac{\mathcal{M}}{(1-\mathcal{M})^{3/5}}\,.\label{dragforceapplate}
\end{equation}
The dependence of the early- and late-time drag force on the Mach number of the perturber is shown in Fig.~\ref{fig:slabsubsonicforcem}. In the subsonic regime, the dynamical friction due to a three-dimensional slab medium with Dirichlet (Neumann) conditions is always smaller (larger) in magnitude than the one due to an infinite three-dimensional medium.

\subsection{Supersonic perturber}
For supersonic-moving perturbers, $\mathcal{M}>1$, the argument of the delta function
in Eq.~\eqref{slabdensity} has roots only if
\begin{equation}
s\le-\sqrt{\left(\mathcal{M}^2-1\right)\left[y^2+\left(z-2 l L\right)^2\right]}\,.
\end{equation}
In that case, those roots are 
\begin{equation}
w_{l,\mp}\equiv\frac{1}{\mathcal{M}^2-1}\left[s \mp \mathcal{M} \sqrt{s^2-(\mathcal{M}^2-1)\left[y^2+\left(z-2 l L\right)^2\right]}\right]\,.\nonumber
\end{equation}
With some algebra one can show that Eq.~\eqref{slabdensity} gives
\begin{widetext}
	\begin{align} 
	&\alpha(\textbf{r},t)=\frac{\mathcal{G} M}{c^2}\sum_{l=-\infty}^{+\infty}  \frac{(-1)^{\eta l}}{\sqrt{s^2-(\mathcal{M}^2-1)\left[y^2+(z- 2l L)^2\right]}} \bigg(\Theta\left[c^2 t^2- x^2-y^2-(z-2l L)^2\right] \nonumber\\
	&+2 \Theta\left[c^2 t^2\left(1-\frac{1}{\mathcal{M}^2}\right)-y^2-(z-2 l L)^2 \right] \Theta\left[x-\sqrt{c^2 t^2-y^2-(z-2 l L)^2}\right]\Theta\left[\mathcal{M} c t -x-\sqrt{\left(\mathcal{M}^2-1\right)\left[y^2+(z-2 l L)^2\right]}\right]\bigg)\, ,
	\end{align}
\end{widetext}
where we are considering the Heaviside function to vanish when evaluated over non-real numbers.
\begin{figure}[]
	\includegraphics[width=1\linewidth]{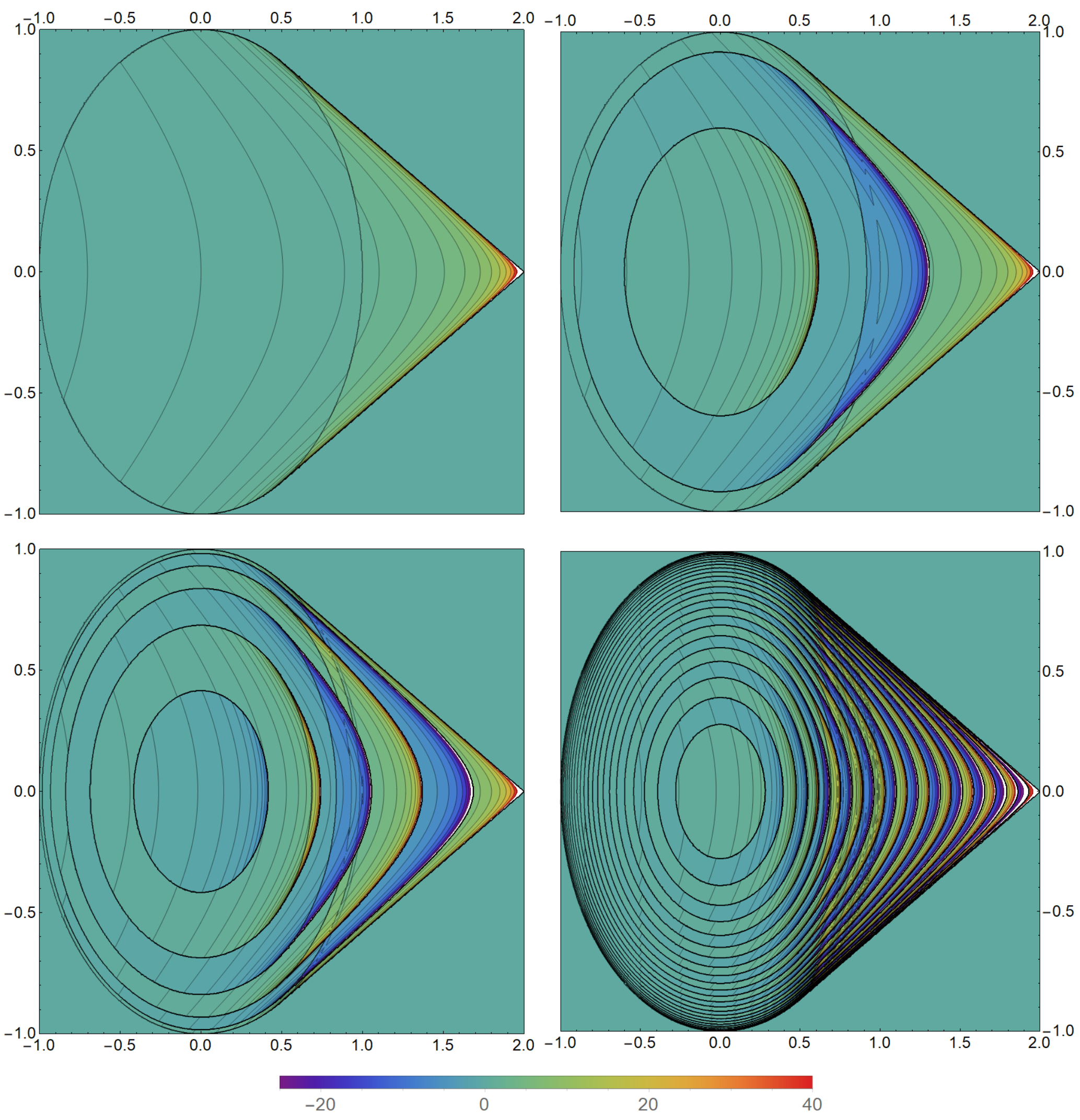}
	\caption{Density perturbation $\alpha c^3 t/(\mathcal{G} M)$ in a gaseous slab, along $z=0$, due to the gravitational interaction with a supersonic perturber with Mach number $\mathcal{M}=2$, for $c t/L=0.5,5,11,50$ (left to right, up to down). The horizontal axis represents the coordinate $x/(c t)$, and the vertical axis the coordinate $y/(c t)$. The contours represent curves of uniform density. The observed ripples are echoes of the original density fluctuation. Each ripple is associated with an $l$-term. At linear order approximation, there is an infinite-density shock wave with conic shape ($l=0$), and shock wave echoes (other $l$-terms) located inside the conic surface.}
	\label{fig:slabsupsonicz0D}
\end{figure}

The perturbation in the gas density, along the $z=0$ plane, caused by a supersonic particle with $\mathcal{M}=2$ is shown in Fig.~\ref{fig:slabsupsonicz0D} at different instants. As expected, for early times $c t/L< 1$, all the results are identical to those in infinite media~\citep{Ostriker:1998fa}~\footnote{Interestingly, the late-time results for the perturbed density profile in a three-dimensional slab with Neumann conditions mimic those obtained in a truly two-dimensional setting (where the gravitational force falls with $\sim 1/r$, instead of the usual $\sim 1/r^2$), in both subsonic and supersonic regimes.}.

In the subsonic regime the density perturbation was infinite only {\it at} the particle location, and surfaces of constant density in the neighborhood of this point were concentric oblate spheroids centered at it, with short-axis along the $x$ direction. Thus, coincidentally, the front-back symmetry of the medium density about the particle suppressed the contribution of this region to the drag force, assuring its finiteness~\citep{Ostriker:1998fa,Rephaeli1980ApJ}. Obviously, this is not the case in the supersonic regime. In fact, it can be shown that the drag force felt by a supersonic point-like particle is infinite.
Thus, a regularization procedure needs to be introduced. We follow the standard, physically motivated, procedure of describing actual sources via an effective size $r_{\text{min}}$. This produces a cutoff in the force integral, describing the effective size of the particle, and assuring that the drag remains finite.

\begin{figure}[]
	\includegraphics[width=1\linewidth]{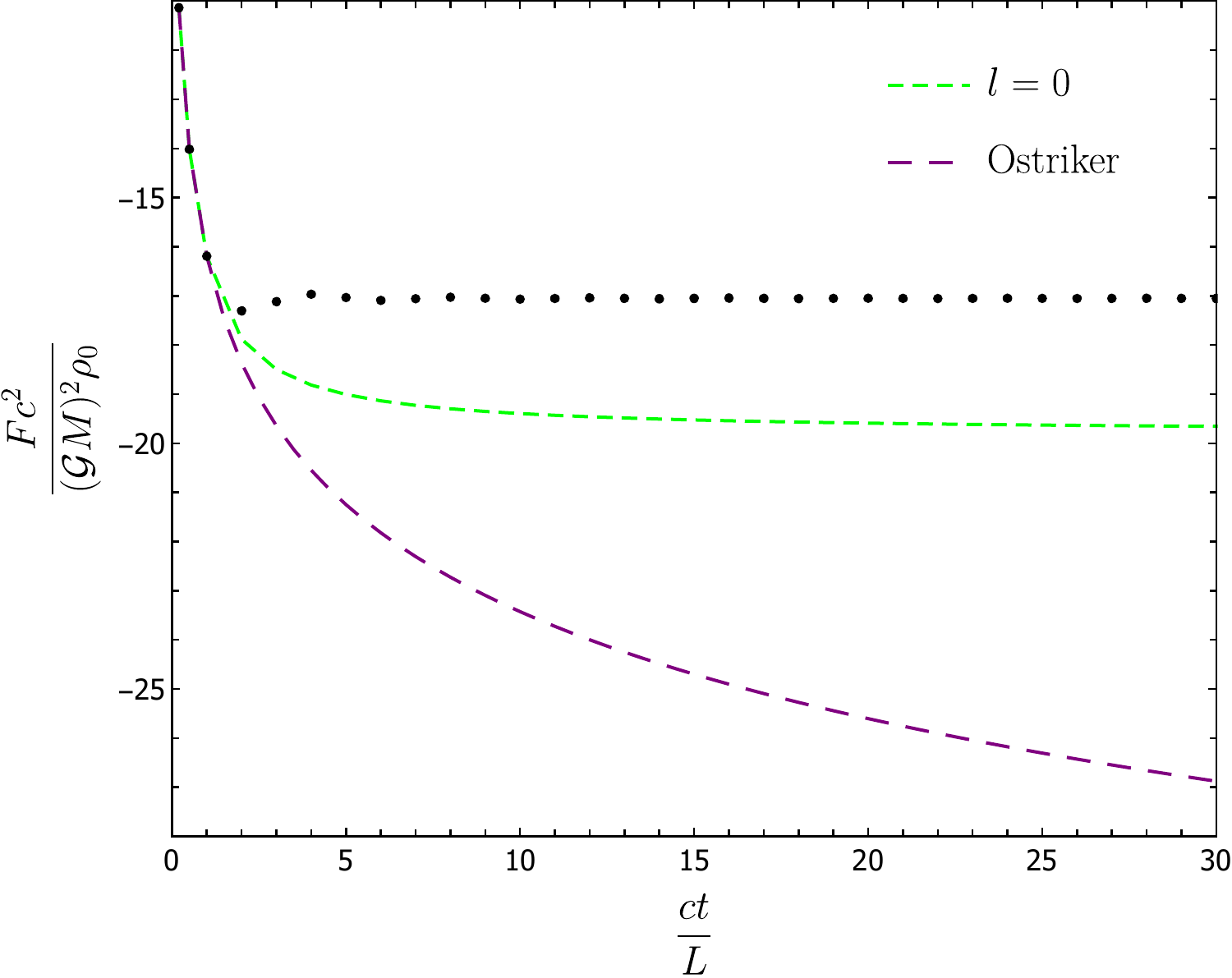} 
	\caption{Time-dependence of the gravitational drag force acting on a supersonic particle with Mach number $\mathcal{M}=2$, and size $r_\text{min}/L=10^{-2}$ (black dots). At early times $c t/L<1$, the dots are in agreement with the drag formula of Eq.~\eqref{supersonic_ostriker} for non-compact mediums (purple dashed curve). At late times $c t/L \gg 1$, our results are well described by Eq.~\eqref{slabDragforcesupappD}. In green, we show the ($l=0$) contribution from the non-reflected wake.}
	\label{fig:slabsupsonicforcetm2D}
\end{figure}

Figure~\ref{fig:slabsupsonicforcetm2D} shows the time-dependence of the drag force, for a fixed Mach number $\mathcal{M}=2$ and $r_\text{min}=10^{-2}L$. At early times $c t/L<1$ we find a drag force identical to that computed in infinite three-dimensional gaseous media~\citep{Ostriker:1998fa}.

\begin{figure}[]
	\includegraphics[width=1\linewidth]{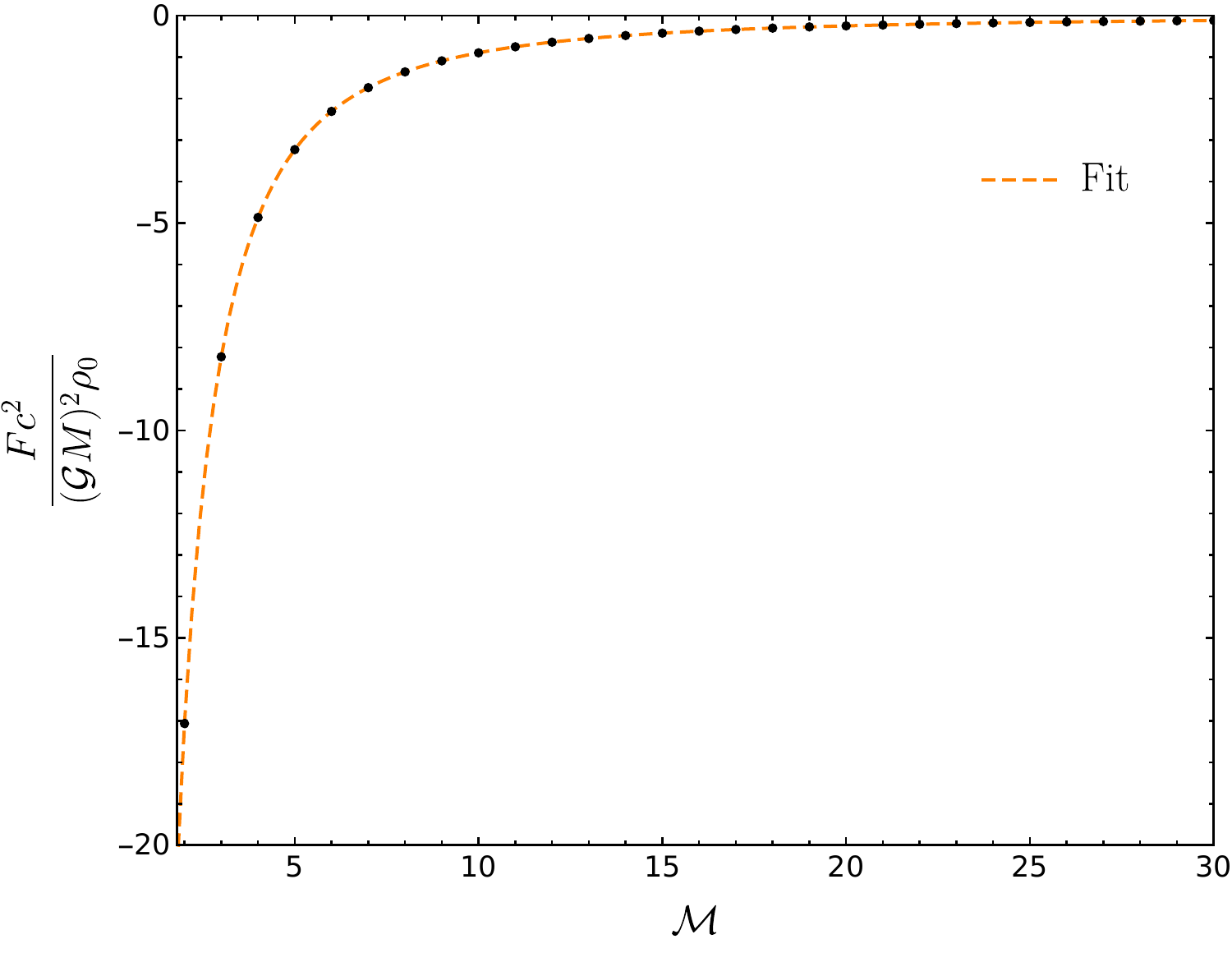} 
	\caption{Late-time drag force for supersonic particles with finite size $r_\text{min}/L=10^{-2}$ (black dots). The results are well approximated by the fit expression~\eqref{slabDragforcesupappD} for $\mathcal{M}>2$ (orange dashed curve).}
	\label{fig:slabsupsonicforcelateD}
\end{figure}

Surprisingly, at late times $c t/L\gg 1$ the gravitational drag felt by a particle moving supersonically through a slab with Dirichlet conditions is time-independent. In fact, our numerical results show (see Fig.~\ref{fig:slabsupsonicforcelateD}) that this late-time drag force is well described by
\begin{equation} \label{slabDragforcesupappD}
F\simeq -\frac{(\mathcal{G} M)^2 \rho_0}{c^2} \left(\mathcal{D}+ \frac{4\pi}{\mathcal{M}^2} \log\left(L/r_\text{min}\right)\right) \,,
\end{equation} 
with $\mathcal{D}\equiv (21.17 \mathcal{M}^{0.83} -22.05)/\mathcal{M}^{2.58}$,
for Mach number $\mathcal{M}>2$. The magnitude of the drag force increases when the size of the particle decreases, but it is a very mild, logarithmic, dependence. For fixed $\mathcal{M}$, the second term in the equation above is dominant for a sufficiently small perturber $L/r_\text{min}\gg 1$.

With Neumann conditions, our numerical results show that, at late times $c t/L \gg 1$, the drag force is well described by
\begin{equation}
F\simeq -\frac{(\mathcal{G} M)^2 \rho_0}{c^2} \left[\mathcal{J}+ \frac{4 \pi}{\mathcal{M}^2} \log\left(\mathcal{M} \frac{c t}{L}\frac{L}{r_\text{min}}\right)\right]\,,
\end{equation}
where $\mathcal{J}\sim 1$ is a function of $\mathcal{M}$.
This is the same late-time ($c t/r_{\text{min}}\gg 1$) behavior of the drag force felt by a particle moving at supersonic speed through a non-compact three-dimensional medium:
\begin{equation}
F\simeq -\frac{4 \pi}{\mathcal{M}^2} \frac{(\mathcal{G} M)^2 \rho_0}{c^2} \log\left(\mathcal{M}\frac{c t}{r_\text{min}}\right)\,,
\end{equation}
which was obtained in Ref.~\cite{Ostriker:1998fa} [see Eq.~\eqref{supersonic_ostriker}]. Thus, interestingly, in a slab with Neumann boundary conditions, both the early- and late-time drag force have the same behavior as in non-compact geometries.  

As it happened in the subsonic regime: the gravitational drag felt by a particle moving at supersonic speed through a three-dimensional slab medium with Dirichlet (Neumann) conditions is always smaller (larger) in magnitude than the one it would feel if moving through an infinite three-dimensional medium.

%

\section{Conclusions and outlook}\label{sec:conclusions}
In this work we computed, for the first time, the gravitational drag force felt by a massive particle moving in a straight line through a three-dimensional slab-like medium, taking into account reflections of the wake on the boundaries. Our results show that the late-time drag force is strongly dependent on the boundary conditions of the slab. Nevertheless, the physically relevant slab-like setups satisfy Dirichlet conditions (on $\alpha$) at the boundaries. In those setups, the drag force can be substantially reduced relatively to extended media. However, since the reflections of the medium wake play an important role in our study, it is important to understand if these reflections are also present in a (realistic) vertically stratified open medium. Otherwise, the conclusions obtained with this simple setup could not be extrapolated to more realistic astrophysical setups. In the Appendix, we show that, indeed, wake reflections are also present in open media, provided that their density falls off sufficiently fast in the vertical direction.

We should highlight that Ref.~\cite{namouni} also studied the time-dependent dynamical friction on compact homogeneous media. In particular, the effect of wake reflections on the boundaries was investigated. However, only one wake reflection was considered, and, though not explicitly stated, Neumann boundary conditions were used. As we explained before, and show in the appendix, the (physically) realistic boundary conditions are of Dirichlet type. Thus, an important conclusion was missed: that, generically, wake reflections tend to suppress gravitational drag.

It is worth pointing out that estimates for the steady-gravitational drag felt by a perturber moving in a straight line through a very thin disk, using a two-dimensional approximation to describe the disk, were made previously~\citep{Muto:2011qv}. This approximation is very good at describing the contribution to gravitational drag coming from fluctuations far from the perturber, which have already felt the slab boundaries. In such setups, the dominant contribution to the subsonic motion drag comes from far regions, and the approximation is expected to hold in that regime~\citep{Muto:2011qv}. For an inviscid medium, the gravitational drag was estimated to be suppressed with $1/t$. This is in very good agreement with our own results (see Eq.~\eqref{fitsubD}). However, the approximation in Ref.~\cite{Muto:2011qv} fails at describing the contribution to the drag from the near region, which is the dominant one in supersonic motion. Nevertheless, though not succeeding in obtaining the correct dependence on $L/r_\text{min}$, they estimate the supersonic late-time drag to be steady and proportional to $1/\mathcal{M}^2$, which is in agreement with our results for $L/r_\text{min}\ll1$ (see Eq.~\eqref{slabDragforcesupappD}). In that case (sufficiently small perturber), we recover the well-known estimates for the steady supersonic drag force in a three-dimensional medium with effective size $L$, both in collisional media~\citep{Dokuchaev1964,Ruderman1971,Rephaeli1980ApJ,canto} and collisionless media~\citep{Tremaine1987}.\footnote{In the case of collisionless media, there is no notion of sound speed. Nevertheless, the analogous regime to the supersonic motion is when the perturber has a velocity much larger than the particle dispersion velocity of the medium~\citep{Ostriker:1998fa}.} Again, this is related to the fact that, for supersonic motion, the dominant contribution to the drag force comes from the near region. So, one does not expect the wake reflections to play an important role in the drag; all the more so for a very small particle.

We expect our results to be important to the study of the physics of accretion and protoplanetary disks. There is a substantial body of theoretical and numerical studies on the disk-planet gravitational interaction \citep{Tremaine79,Tremaine80,Ward86,Tanaka_2002,Muto:2011qv,Stone2018}. However, in most of them two oversimplifications are used: (i) the disks are assumed to be very thin, and a two-dimensional approximation is used to treat the medium; (ii) the gravitational wake is assumed to be completely dissipated at the boundaries, without any reflection. A full three-dimensional treatment of the gravitational interaction between a planet and a disk, not assuming (i), but maintaining assumption (ii) finds the following~\citep{Tanaka_2002}: the migration time of an Earth-sized planet at $5$AU is of the order of $10^6$yr, which is $2$ or $3$ times longer than previously obtained results using the two-dimensional approximation~\citep{Hayashi}. Their result is very relevant: since the formation time of a giant planet at $5$AU is of the order of $10^6$yr \citep{Tanaka}, the planetary migration must happen in a longer, or, at least, comparable, timescale, to explain the existence of giant planets. In that same work, they also suggested that the reflection of the gravitational wake on the disk edges, which they neglected, could weaken even more the disk-planet interaction, and increase the planet migration time. Our results clearly support their intuition in the case of subsonic motion, where the drag force is strongly suppressed (see Fig.~\ref{fig:slabsubsonicforcetD}). For supersonic motion, the $l=0$ term, which is not sensitive to the boundaries, accounts for most of the late-time gravitational drag. Thus, even though the drag force is also suppressed in the case of supersonic motion, we do not expect the effect of wake reflections to be as striking as in the subsonic case. 

One can argue that all the results derived here assume linear motion and cannot, formally, be applied in setups involving circular motion. Despite this being true, Ref.~\cite{Kim:2007zb} obtained the remarkable result that the drag force formulae derived for linear motion in extended media by Ref.~\cite{Ostriker:1998fa} give reasonably good estimates for the drag felt by circular-orbit perturbers. We expect the same thing to happen here, at least qualitatively. In fact, the approach of Ref.~\cite{Kim:2007zb,Kim:2008ab} to extend the drag formulae derived in Ref.~\cite{Ostriker:1998fa} from linear motion to circular-orbit and binary motion, respectively, can, in principle, be applied in a straightforward way to extend our results to those same motions.

The unbounded-medium approximation derived in Ref.~\cite{Kim:2007zb} was used recently to estimate the impact of dynamical friction in thin accretion disks on gravitational-wave observables~\citep{Barausse:2014tra}. It was concluded that dynamical friction may indeed be important and lead to degradation of gravitational-wave templates for detection. Notice that, in a physically realistic setup, the accretion disk height is $L\sim  r c/v_K$, where $r$ is the distance from the disk center, and $v_k\equiv (\mathcal{G} M/r)^{1/2}$ is the local Keplerian velocity at which the perturber is moving in its circular-orbit motion. In general, both $c$ and $L$ are function of $r$. Nevertheless, Ref.~\cite{Barausse:2014tra} assume that the relative velocity of the perturber with respect to the disk is $\mathcal{M}\simeq v_k/c \sim r/L$, and, so, for a thin accretion disk $r/L\gg 1$, the motion is supersonic. So, from what we discussed above, the dominant contribution to the drag force comes from the region near the perturber. Thus, even though our toy model neglects variations of $c$ and $L$, we still expect it to describe appropriately the present setup. Moreover, the sound travel time to the disk edges is of the same order of the orbital-motion period (\textit{i.e.}, $c/L \sim v_k/r$). Thus, the finiteness effects of the disk may be relevant for the gravitational drag force in thin accretion disks, and can, possibly, change the conclusion of~Ref.~\cite{Barausse:2014tra}.

\begin{acknowledgements}
	
R.V.\ was supported by the FCT PhD scholarship SFRH/BD/128834/2017.
V.C.\ acknowledges financial support provided under the European Union's H2020 ERC 
Consolidator Grant ``Matter and strong-field gravity: New frontiers in Einstein's 
theory'' grant agreement no. MaGRaTh--646597.
M.Z.\ acknowledges financial support provided by FCT/Portugal through the IF
programme, grant IF/00729/2015.
This project has received funding from the European Union's Horizon 2020 research and innovation programme under the Marie Sklodowska-Curie grant agreement No 690904.
We acknowledge financial support provided by FCT/Portugal through grant PTDC/MAT-APL/30043/2017.
The authors would like to acknowledge networking support by the GWverse COST Action 
CA16104, ``Black holes, gravitational waves and fundamental physics.''

\end{acknowledgements}
\appendix*
\section{Wake reflections in open media} \label{sec:app}
Here, we show that the wake reflections observed in this work are not specific to unphysical slab media with Dirichlet boundary conditions. Instead, they are also present in realistic vertically stratified (open) media, provided that their density falls sufficiently fast to zero. Moreover, we show that (as far as wake reflections is concerned) these stratified setups are well modeled by a slab with Dirichlet conditions at the (cutoff) boundaries.

Let us start by considering a vertically stratified isothermal gaseous medium with unperturbed density $\rho_0(z)$. Here, we focus on the $z$-direction dynamics. Thus, the linearized equation describing the relative perturbed density is
\begin{equation} \label{EOMstrat}
\frac{\partial^2}{\partial z^2} \alpha-\frac{1}{c^2}\frac{\partial^2}{\partial t^2}\alpha + \left(\frac{\partial}{\partial z} \log \rho_0\right) \frac{\partial}{\partial z}\alpha=0\,.
\end{equation}
Defining $\bar{\alpha}\equiv \alpha(z,t) e^{-k(z)}$ with
\begin{equation*}
k\equiv-\frac{1}{2}\log \left(\frac{\rho_0(z)}{\rho_0(0)}\right)\,,
\end{equation*}
Eq.~\eqref{EOMstrat} gives
\begin{equation} \label{EOMstrat2}
\frac{\partial^2}{\partial z^2} \bar{\alpha}-\frac{1}{c^2}\frac{\partial^2}{\partial t^2}\bar{\alpha}+ \left[k''-\left(k'\right)^2\right]\bar{\alpha}=0\,,
\end{equation}
where $k'$ denotes the derivative of $k$ with respect to $z$. Now, one can write $\bar{\alpha}$ as the Fourier-integral 
\begin{equation}
	\bar{\alpha}=\int d\omega\, \bar{\alpha}_\omega(z) e^{-i \omega t} \,,
\end{equation} 
which after substitution in Eq.~\eqref{EOMstrat2} gives
\begin{equation}
	\frac{\partial^2}{\partial z^2} \bar{\alpha}_\omega+q_\omega(z) \bar{\alpha}_\omega=0\,,
\end{equation}
with
\begin{equation}
	q_\omega \equiv \left(\frac{\omega}{c}\right)^2+k''-\left(k'\right)^2\,.
\end{equation}
By looking at the sign of $q_\omega$, one can identify the regions where $\omega$-mode fluctuations of the medium density propagate, and the ones where they evanesce: propagation happens in regions with positive $q_\omega$, and evanescence in regions with negative $q_\omega$ \citep{Kumar}.

As an example, consider the unperturbed density profile
\begin{equation}\label{expsetup}
	\rho_0=\rho_0(0)\left[1+(e^{-z/h}-1)\Theta(z)\right]\,,
\end{equation}
with $h$ the effective thickness of the medium's edge, and $\Theta(z)$ the Heaviside step function. This density profile gives
\begin{equation}
	q_\omega=\left(\frac{\omega}{c}\right)^2-\Theta(z) \left(\frac{1}{2 h}\right)^2\,.
\end{equation}
We see that every $\omega$-mode can propagate in $z<0$, whereas only the $|\omega|>c/(2 h)$ modes can propagate in $z>0$. In other words: an $\omega$-mode coming from $z<0$, and propagating in the positive $z$-direction, gets totally reflected at $z=0$, iff $|\omega|\leq c/(2 h)$; otherwise, the $\omega$-mode is partially reflected and partially transmitted. The frequency $\omega=c/(2 h)$ is often called cutoff frequency \citep{Lamb}.

The gravitational wake produced by a moving perturber can be modeled by a real wave packet with spatial width $\delta z \sim 2 L$, where $2 L$ is the effective thickness of the medium. Thus, the Fourier-transform in space of this gravitational wake is centered at $\omega/c=0$, and has width $\delta \omega /c \sim 1/(2\delta z)\sim 1/(4L)$.\footnote{This can be seen through an uncertainty principle for Fourier transformations, assuming a Gaussian-like wave packet.} So, the typical frequency content of a gravitational wake produced in slab-like media has $\delta \omega \sim c/(4 L)$. Thus, if $h\ll 2L$, the wake is totally reflected at $z=0$. In that case, concerning the wake reflections, this stratified medium is well modeled by an homogeneous medium ($z<h$) with Dirichlet conditions at the $z=h$ (cutoff) boundary.\footnote{Although there is no propagation at $z>0$, the stratified edge introduces a phase shift in the wave packet. Thus, in order to take this effect into account, we choose $z=h$ as the cutoff boundary, instead of $z=0$.} 

\begin{figure}[]
	\begin{tabular}{c}
		\includegraphics[width=0.95\linewidth]{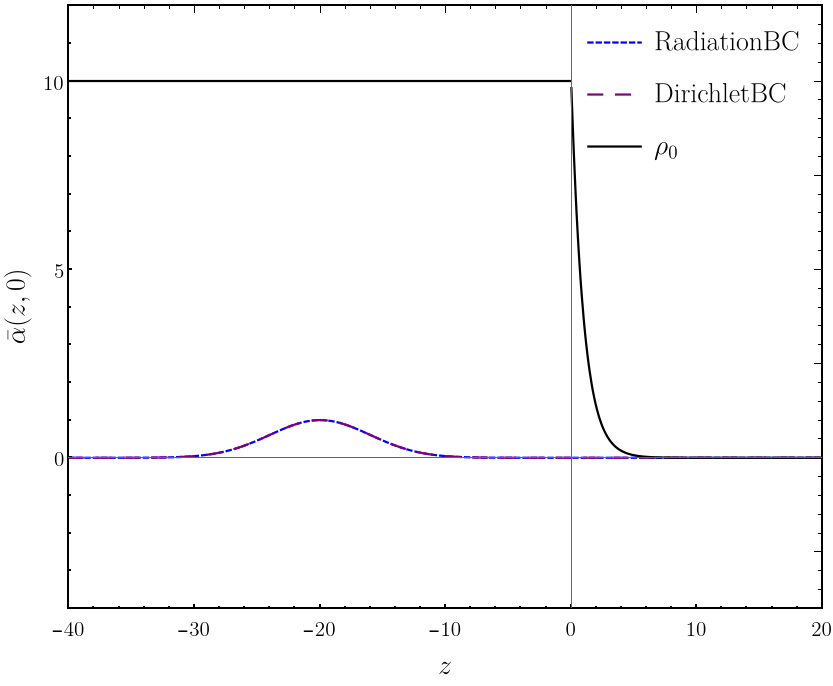} \\
		\includegraphics[width=0.95\linewidth]{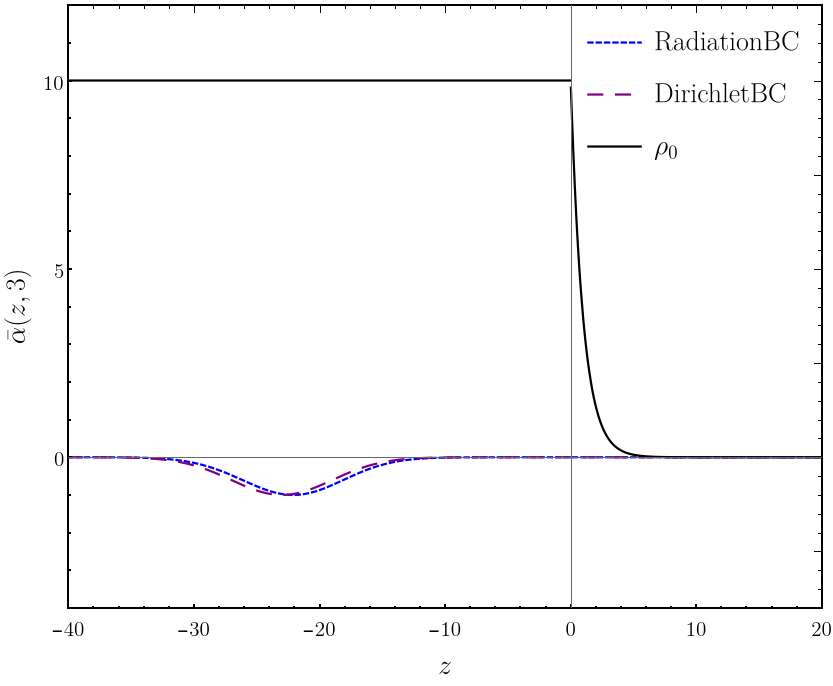} 
	\end{tabular}
	\caption{Time-evolution of a wave packet with initial conditions $\bar{\alpha}(z,0)=e^{-\frac{1}{2}\left(\frac{z+20}{4}\right)^2}$ and $\partial_t \bar{\alpha}(z,0)=-c\, \partial_z \bar{\alpha}(z,0)$ by Eq.~\eqref{EOMstrat2}. \textbf{Blue:} wave packet propagating in the stratified setup~\eqref{expsetup} (also represented in the figure), with radiation boundary conditions; \textbf{Purple:} wave packet propagating in an homogeneous medium with Dirichlet conditions at $z=h$. \textbf{Above:} initial (incident) wave packets propagating from left to right; \textbf{Below:} reflected wave packets propagating from right to left. 
	The parameters used were: $c=15$, $\rho_0(0)=10$, and $h=1$.}
	\label{fig:expedge}
\end{figure}

Figure~\ref{fig:expedge} shows the results for the time-evolutions of: a wave packet propagating in the stratified setup~\eqref{expsetup}, with (open) radiation boundary conditions; and, a wave packet propagating in an homogeneous medium with Dirichlet conditions at $z=h$. These results are in accordance with the predictions above.

Finally, let us consider an additional example. For a disk edge in isothermal equilibrium, the unperturbed density is \citep{shakura_sunyaev}
\begin{equation}\label{gausssetup}
\rho_0=\rho_0(0)\left[1+\left(e^{-\frac{1}{2}\left(\frac{z}{h}\right)^2}-1\right)\Theta(z)\right]\,.
\end{equation}
This profile gives
\begin{equation}
q_\omega=\left(\frac{\omega}{c}\right)^2-\Theta(z) \left(\frac{z^2}{4 h^4}- \frac{1}{2h^2}\right)\,.
\end{equation}
We see that each $\omega$-mode can only propagate in the region
\begin{equation*}
	z<z_\omega \equiv h \sqrt{2+\left(\frac{2 h \omega}{c}\right)^2}\,,
\end{equation*}
being evanescent elsewhere. Moreover, since the typical frequency content of a gravitational wake produced in slab-like media has $\delta \omega \sim c/(4L)$; if the edges are sufficiently thin ($h\ll 2 L$), then $z_\omega \sim \sqrt{2} h$, for all frequencies composing the wave packet. In other words: the whole packet is totally reflected at $z=\sqrt{2}\,h$. Thus, again, concerning the wake reflections, this stratified medium is well modeled by an homogeneous medium ($z<h$) with Dirichlet conditions at the $z=h$ (cutoff) boundary.\footnote{As in the last example, we chose $z=h$ as the cutoff boundary, instead of $z=\sqrt{2} h$, in order to model the correct phase shift introduced by the reflection in the stratified medium.}

\begin{figure}[]
	\begin{tabular}{c}
		\includegraphics[width=0.95\linewidth]{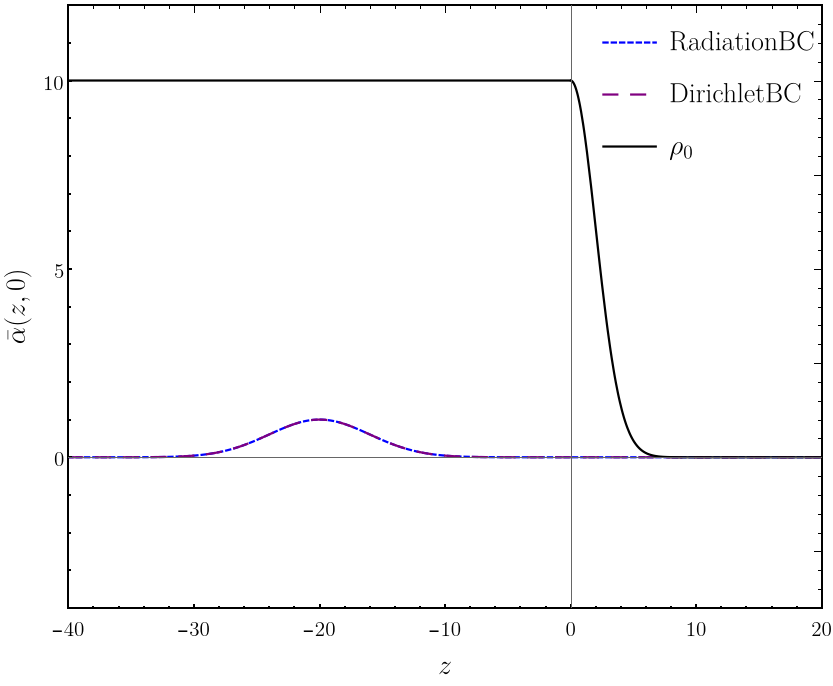} \\
		\includegraphics[width=0.95\linewidth]{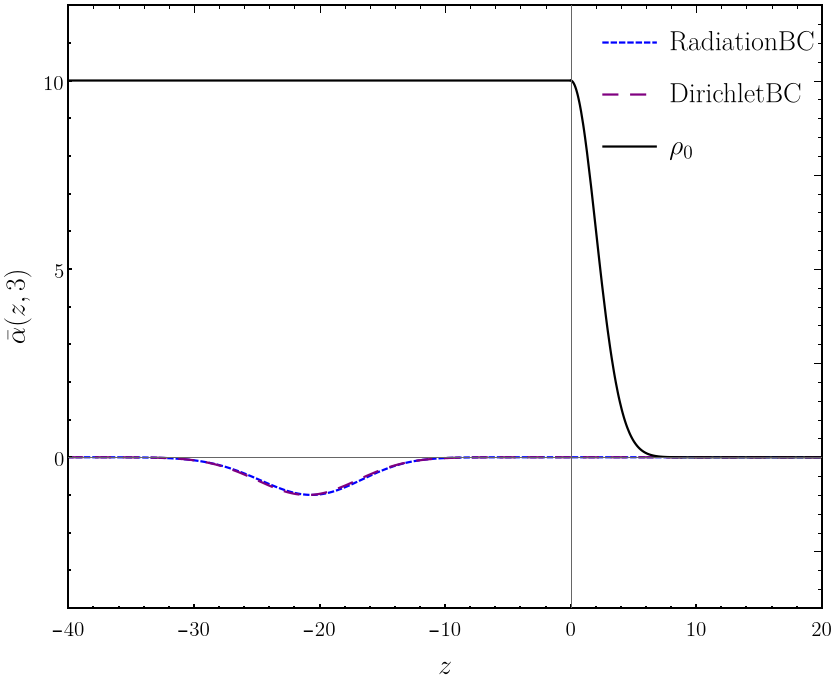} 
	\end{tabular}
	\caption{Time-evolution of a wave packet with initial conditions $\bar{\alpha}(z,0)=e^{-\frac{1}{2}\left(\frac{z+20}{4}\right)^2}$ and $\partial_t \bar{\alpha}(z,0)=-c\, \partial_z \bar{\alpha}(z,0)$ by Eq.~\eqref{EOMstrat2}. \textbf{Blue:} wave packet propagating in the stratified setup~\eqref{gausssetup} (also represented in the figure), with radiation boundary conditions; \textbf{Purple:} wave packet propagating in an homogeneous medium with Dirichlet conditions at $z=h$. \textbf{Above:} initial (incident) wave packets propagating from left to right; \textbf{Below:} reflected wave packets propagating from right to left. 
		The parameters used were: $c=15$, $\rho_0(0)=10$, and $h=2$.}
	\label{fig:gaussedge}
\end{figure}

Figure~\ref{fig:gaussedge} shows the results for the time-evolutions of: a wave packet propagating in the stratified setup~\eqref{gausssetup}, with (open) radiation boundary conditions; and, a wave packet propagating in an homogeneous medium with Dirichlet conditions at $z=h$. These results are again in accordance with the predictions above.

As a final note, we point out that, in this appendix, we also show that the boundary conditions that a physically realistic slab medium satisfies are the Dirichlet (reflection with inversion) ones. Had we used Neumann (reflection without inversion) conditions for the time-evolutions, the reflected wave packets would be inverted with respect to the wave packets reflected by the (realistic) stratified media.

\clearpage
\bibliographystyle{apsrev4}
%


\end{document}